\documentclass[12pt,twoside]{article}
\usepackage{wrapfig}
\usepackage{graphicx}
\usepackage{cmp2e}
\title[A microscopic theory of photonucleation:
Density functional approach]
{A microscopic theory of photonucleation:
Density functional approach for the properties of a fluid of two-level atoms,
\\
a part of which is excited}
\author[O.~Derzhko and V.~Myhal]{Oleg~Derzhko\refaddr{label1} and Vasyl~Myhal\refaddr{label2}}
\addresses{
\addr{label1} Institute for Condensed Matter Physics NASU,\\
              1 Svientsitskii Street, 79011 L'viv, Ukraine
\addr{label2} The Ivan Franko National University of L'viv,\\
              Department for Theoretical Physics,\\
              12 Drahomanov Street, 79005 L'viv, Ukraine}
\begin{document}

\maketitle

\begin{abstract}
We use the density functional method 
to examine the properties of the nonuniform (two-phase) fluid of two-level atoms,
a part of which is excited.
From the analysis of the equation of state of a gas of two-level atoms,
a part of which is excited,
the following density functional of the grand thermodynamical potential emerges
\begin{eqnarray}
\Omega[\rho({\bf{r}})]
=\Omega_{{\rm{CS}}}[\rho({\bf{r}})]
-\frac{6\sigma^3a(c_1,T)}{\pi}
\int_{\vert{\bf{r}}_1-{\bf{r}}_2\vert\ge 2\sigma}{\rm{d}}{\bf{r}}_1{\rm{d}}{\bf{r}}_2
\frac{\rho({\bf{r}}_1)\rho({\bf{r}}_2)}{\vert{\bf{r}}_1-{\bf{r}}_2\vert^6}
\nonumber
\end{eqnarray}
with
\begin{eqnarray}
a(c_1,T)
=\frac{1}{32}a^2v(E_1-E_0)
\left(c_0-c_1+2c_0c_1\frac{E_1-E_0}{kT}\right)
\nonumber
\end{eqnarray}
(here $\Omega_{{\rm{CS}}}[\rho({\bf{r}})]$ is the Carnahan-Starling term, 
$\sigma$ is the atom radius,
$v=\frac{4}{3}\pi\sigma^3$, 
$c_1$ is the concentration of excited atoms,
$c_0+c_1=1$,
$E_1-E_0$ is the excitation energy
and $a$ is the dimensionless parameter 
which characterizes the atom).
We use this expression to calculate the nucleation barrier for vapor-to-liquid phase transition 
in the presence of excited atoms.
\pacs 64.70.Fx, 82.65.Dp, 62.60.Nh, 64.60.Qb
\end{abstract}

The studies of the equilibrium properties 
of a gas of identical atoms,
a part of which is in an excited electronic state, 
attract attention during last forty years
\cite{01,02,03,04,05,06,07,08,09,10}.
Such atoms may appear owing to the electromagnetic irradiation with the frequency,
which corresponds to the excitation energy of the atom.
Since the life-time of the excited state 
is essentially larger than the time required for establishing of equilibrium 
over translational degrees of freedom,
the system should exhibit equilibrium properties 
at a given (nonequilibrium) concentration of excited atoms.
Moreover,
owing to new effective long-range interatomic interactions
-- the resonance dipole-dipole interactions --
one may expect essential changes 
of various equilibrium characteristics due to the presence of excited atoms.
The analysis performed within the frames of the cluster expansion method
confirmed these expectations
(see Refs. \cite{07,08} and references therein).
On the other hand,
there are only a few papers 
which may be related to experimental observations 
of the theoretically investigated features of the gas with excited atoms.
We should mention here the papers on the influence of irradiation 
on the condensation of iodine and anthracene vapor
\cite{11,12}
and 
on the photonucleation 
\cite{13,14,15,16,17,17a,17b,17c,17d},
in particular,
in vapors of mercury and cesium.
The latter studies reported 
the quantitative results of resonance irradiation influence 
on the nucleation rate. 
The main conclusion of these studies is as follows:
the resonance irradiation of the nucleation zone leads to a sharp increase of the nucleation rate.
The obtained experimentally results,
to our best knowledge,
have not been explained until now.
One may try to interpret these data basing on the theory of nucleation 
in a supersaturated vapor which contains excited atoms.
Such atoms may appear as a result of the resonance irradiation.

In what follows we present preliminary results about the nucleation phenomena 
in a fluid of two-level atoms, 
a part of which is excited
(see also \cite{18}).
In our study 
we use the density functional approach developed by D.~W.~Oxtoby with coworkers \cite{19}.
This method permits to obtain the nucleation rate 
basing on the first principles.
The nucleation rate $J$ is connected with the nucleation barrier $A$,
$J=J_0\exp\left(-\frac{A}{kT}\right)$.
The nucleation barrier can be calculated 
within the frames of the classical nucleation theory 
which relies on the capillarity approximation \cite{19}
\begin{eqnarray}
\frac{A^{\rm{cl}}}{kT}
=\frac{16\pi}{3}\left(\frac{\gamma}{kT}\right)^3\frac{1}{\rho_l^2\ln^2s}.
\label{01}
\end{eqnarray}
Here $\gamma$ is the surface tension of the vapor-liquid interface,
$\rho_l$ is the density of liquid,
$s=\frac{p}{p_0}$ is the supersaturation,
$p$ is the actual pressure of the supersaturated vapor
and $p_0$ is the equilibrium pressure.
To compute the nucleation rate according to Eq. (\ref{01})
one has at first to construct the vapor-liquid phase diagram
determining the equilibrium pressure $p_0$ and the liquid density $\rho_l$ 
at the temperature $T$
and then to compute the surface tension $\gamma$ at this temperature.
This calculation can be done within the density functional approach 
considering the planar vapor-liquid interface. 
Alternative approach for calculation of the nucleation barrier 
which does not use the key assumption in the classical nucleation theory 
-- the capillarity approximation --
was suggested by D.~W.~Oxtoby \cite{19}.
According to this scheme 
one has to consider a metastable vapor 
in a spherical vessel of the radius ${\cal{R}}$,
assume appearance of a spherical liquid droplet in the center of the vessel 
and analyze the density profile 
and the value of the grand thermodynamical potential 
of such a two-phase fluid with the spherical vapor-liquid interface.
The value of the grand thermodynamical potential of such a metastable fluid 
$\Omega(T,\mu,V)$ permits to calculate the nucleation barrier
via the equation
\begin{eqnarray}
A=\Omega(T,\mu,V)-\left(-p\frac{4}{3}\pi{\cal{R}}^3\right).
\label{02}
\end{eqnarray}
In our study we use both schemes for calculation 
of the vapor-to-liquid nucleation barrier
in a system of two-level atoms,
a part of which is excited. 

To perform the theoretical analysis of the vapor-to-liquid nucleation 
in the presence of the excited atoms
we need an appropriate density functional of the grand thermodynamical potential.
We assume the grand thermodynamical potential to be a functional of the density with the form
\begin{eqnarray}
\Omega[\rho({\bf{r}})]
=
kT\int{\rm{d}}{\bf{r}}_1\rho({\bf{r}}_1)
\left(
\ln\left(\Lambda^3\rho({\bf{r}}_1)\right)
+\frac{-1+6v\rho({\bf{r}}_1)-4v^2\rho^2({\bf{r}}_1)}{\left(1-v\rho({\bf{r}}_1)\right)^2}
\right)
\nonumber\\
-\frac{6\sigma^3a(c_1,T)}{\pi}\int_{\vert{\bf{r}}_1-{\bf{r}}_2\vert\ge 2\sigma}
{\rm{d}}{\bf{r}}_1{\rm{d}}{\bf{r}}_2
\frac{\rho({\bf{r}}_1)\rho({\bf{r}}_2)}
{\left\vert {\bf{r}}_1-{\bf{r}}_2 \right\vert^6}
-\mu\int{\rm{d}}{\bf{r}}_1\rho({\bf{r}}_1)
\label{03}
\end{eqnarray}
where
\begin{eqnarray}
a(c_1,T)
=\frac{a^2}{32}v(E_1-E_0)\left(1-2c_1+2(1-c_1)c_1\frac{E_1-E_0}{kT}\right).
\label{04}
\end{eqnarray}
Here $\Lambda$ is the thermal de Broglie wavelength of the atom,
$v=\frac{4}{3}\pi\sigma^3$,
$\sigma$ is the radius of the atom,
$c_1$ is the concentration of the excited atoms,
$a=\frac{d^2/\sigma^3}{E_1-E_0}$
is the dimensionless parameter which characterizes the two-level atom
(in what follows we set $a=1$),
$E_1-E_0$ is the excitation energy,
$d$ is the value of the transitional electrical dipole moment between the ground and excited states.
The equilibrium density minimizes the grand thermodynamical potential 
$\Omega[\rho({\bf{r}})]$,
i.e. is the solution of the following integral equation
\begin{eqnarray}
kT\ln\left(\Lambda^3\rho({\bf{r}}_1)\right)
+kT\frac{8v\rho({\bf{r}}_1)-9v^2\rho^2({\bf{r}}_1)+3v^3\rho^3({\bf{r}}_1)}
{\left(1-v\rho({\bf{r}}_1)\right)^3}
\nonumber\\
-\frac{12\sigma^3a(c_1,T)}{\pi}
\int_{\vert{\bf{r}}_1-{\bf{r}}_2\vert\ge 2\sigma}
{\rm{d}}{\bf{r}}_2
\frac{\rho({\bf{r}}_2)}
{\left\vert {\bf{r}}_1-{\bf{r}}_2 \right\vert^6}
-\mu=0.
\label{05}
\end{eqnarray}
Substituting the equilibrium density into Eq. (\ref{03}) 
one obtains the value of the grand thermodynamical potential of the system $\Omega(T,\mu,V)$.
The adopted density functional of the grand thermodynamical potential (\ref{03}) 
is consistent with the virial state equation obtained earlier \cite{01}.
It takes into account 
the short-range interaction within the Carnahan-Starling local approximation
(the first term in the r.h.s. of Eq. (\ref{03}))
neglecting the difference of the atom radii in the ground and excited states.
Moreover,
it takes into account the long-range interactions,
in particular, 
the resonance dipole-dipole interactions, 
within the mean-field approximation 
(the second term in the r.h.s. of Eq. (\ref{03})).
Note,
that the coefficient $a(c_1,T)$ (\ref{04}) depends on temperature 
only when $c_1$ deviates from zero.
More sophisticated density functionals are available 
but they have not been employed in the present study.

It is convenient to introduce the dimensionless units 
of energy, temperature, chemical potential, length, volume, density, pressure, surface tension etc
renormalizing these quantities as follows:
$E\to \frac{E}{E_1-E_0}$,
$T\to \frac{kT}{E_1-E_0}$,
$\mu\to \frac{\mu}{E_1-E_0}$,
$r\to \frac{r}{\sigma}$,
$V\to\frac{V}{v}$,
$\rho\to v\rho$,
$p\to \frac{pv}{E_1-E_0}$ 
$\gamma\to\frac{\gamma\sigma^2}{E_1-E_0}$ etc,
respectively.
As a result Eqs. (\ref{03}), (\ref{04}), (\ref{05}) become
\begin{eqnarray}
\Omega[\rho({\bf{r}})]
=
\frac{3T}{4\pi}\int{\rm{d}}{\bf{r}}_1\rho({\bf{r}}_1)
\left(
\ln\left(\frac{\Lambda^3}{v}\rho({\bf{r}}_1)\right)
+\frac{-1+6\rho({\bf{r}}_1)-4\rho^2({\bf{r}}_1)}{\left(1-\rho({\bf{r}}_1)\right)^2}
\right)
\nonumber\\
-\frac{9\alpha(c_1,T)}{2\pi^2}\int_{\vert{\bf{r}}_1-{\bf{r}}_2\vert\ge 2}
{\rm{d}}{\bf{r}}_1{\rm{d}}{\bf{r}}_2
\frac{\rho({\bf{r}}_1)\rho({\bf{r}}_2)}
{\left\vert {\bf{r}}_1-{\bf{r}}_2 \right\vert^6}
-\frac{3\mu}{4\pi}\int{\rm{d}}{\bf{r}}_1\rho({\bf{r}}_1),
\label{06}
\end{eqnarray}
\begin{eqnarray}
\alpha(c_1,T)
=\frac{a^2}{32}\left(1-2c_1+\frac{2(1-c_1)c_1}{T}\right),
\label{07}
\end{eqnarray}
\begin{eqnarray}
T\ln\left(\frac{\Lambda^3}{v}\rho({\bf{r}}_1)\right)
+T\frac{8\rho({\bf{r}}_1)-9\rho^2({\bf{r}}_1)+3\rho^3({\bf{r}}_1)}
{\left(1-\rho({\bf{r}}_1)\right)^3}
\nonumber\\
-\frac{12\alpha(c_1,T)}{\pi}
\int_{\vert{\bf{r}}_1-{\bf{r}}_2\vert\ge 2}
{\rm{d}}{\bf{r}}_2
\frac{\rho({\bf{r}}_2)}
{\left\vert {\bf{r}}_1-{\bf{r}}_2 \right\vert^6}
-\mu=0.
\label{08}
\end{eqnarray}
Moreover,
we assume for concreteness in Eqs. (\ref{04}) and (\ref{07}) $a=1$.
We also set without loss of generality $\frac{\Lambda^3}{v}=1$.

\begin{figure}[h]
\vspace{0cm}
\begin{center}
\includegraphics[width=7.0cm]{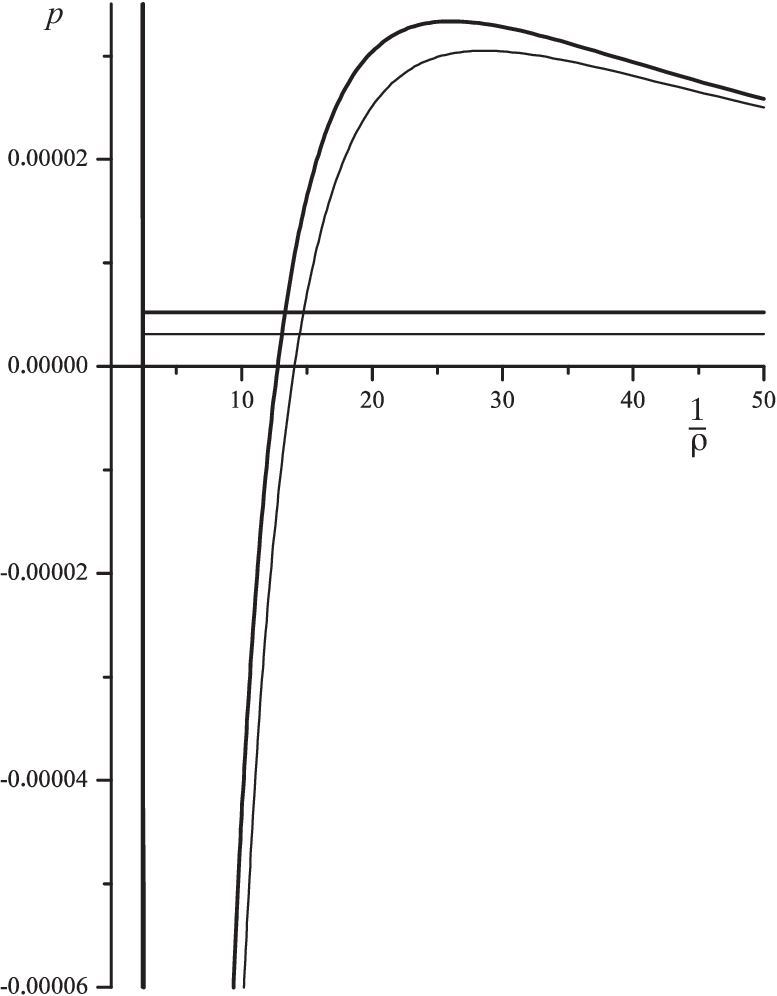}
\hspace{5mm}
\includegraphics[width=7.0cm]{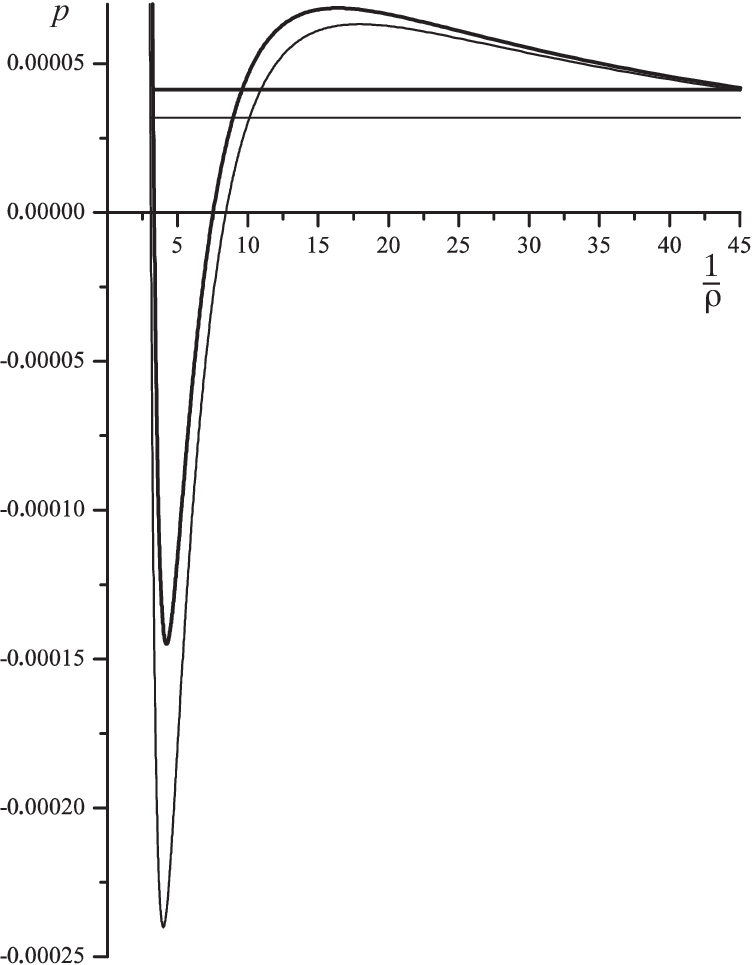}
\end{center}
\caption[]{
The isotherms $p$ vs $\frac{1}{\rho}$ 
at 
$T=0.6T_c(0)$ (left panel)
and
$T=0.8T_c(0)$ (right panel).
Bold curves correspond to $c_1=0$,
thin curves correspond to $c_1=0.00006\;(=0.006\%)$.
The left (right) endpoint of the horizontal part of the isotherm
(corresponding to $p_0$)
gives the liquid density $\rho_l$
(the vapor density $\rho_v$).}
\label{fig1}
\end{figure}

We start with the phase diagram of the system.
For this purpose we assume 
the constancy of the density, 
$\rho({\bf{r}})=\rho$,
that immediately yields instead of Eqs. (\ref{06}) and (\ref{08})
the following expressions  
\begin{eqnarray}
\Omega(\rho)=
T\rho V
\left(
\ln\rho+\frac{-1+6\rho-4\rho^2}{\left(1-\rho\right)^2}
\right)
-\alpha(c_1,T)\rho^2 V-\mu\rho V
\label{09}
\end{eqnarray}
and
\begin{eqnarray}
T\ln\rho+T\frac{8\rho-9\rho^2+3\rho^3}{\left(1-\rho\right)^3}
-2\alpha(c_1,T)\rho-\mu=0.
\label{10}
\end{eqnarray}
Solving Eq. (\ref{10}) 
with respect to $\rho$ and substituting this density into Eq. (\ref{09}) 
one gets the value of the grand thermodynamical potential $\Omega(T,\mu,V)$.
One can also eliminate with the help of Eq. (\ref{10}) the chemical potential $\mu$ 
from Eq. (\ref{09}) getting as a result the equation of state
\begin{eqnarray}
-\frac{\Omega(T,\rho,V)}{TV}
=\frac{p}{T}
=
\rho\frac{1+\rho+\rho^2-\rho^3}{\left(1-\rho\right)^3}
-\frac{a(c_1,T)}{T}\rho^2.
\label{11}
\end{eqnarray}
Eq. (\ref{11}) agrees with the virial state equation 
of a gas of two-level atoms, a part of which is excited, 
obtained earlier \cite{01}
(see also Refs. \cite{07,08}).
Eq. (\ref{10}) may have more than one solution  
which yield the same value of the grand thermodynamical potential $\Omega(T,\mu,V)$.
Indeed, for a given temperature $T$ 
let us fix the value of the grand thermodynamical potential $-\frac{\Omega}{V}=p$
and solve Eq. (\ref{11}) with respect to $\rho$. 
At high temperatures  
(above the critical temperature $T_c$)
one finds only one solution $\rho$
which corresponds to a certain value of the chemical potential $\mu$ in Eq. (\ref{10}).
At low temperatures 
(below the critical temperature $T_c$)
one finds several solutions $\rho$ 
with the corresponding values of chemical potential $\mu$
which follow from Eq. (\ref{10}).
Varying the value of the grand thermodynamical potential $-\frac{\Omega}{V}=p$ 
one finds such two densities $\rho_v$ and $\rho_l>\rho_v$
which yield the same value of $\mu$.
The quantities $T$, $p=p_0$, $\rho_v$, $\rho_l$, $\mu=\mu_0$ 
correspond to the points on the phase diagram 
where the two phases, 
liquid and vapor,
coexist
(see Fig. \ref{fig1}).
In Fig. \ref{fig1}
we display the isotherms which correspond to the temperatures 
$T=0.6T_c(0)\approx 0.00176866$
(left panel)
and
$T=0.8T_c(0)\approx 0.00235822$ 
(right panel)
(here $T_c(0)$ denotes the critical temperature $T_c$ without excited atoms,
i.e. when $c_1=0$)
for two concentrations of excited atoms,
$c_1=0$ (bold curves)
and
$c_1=0.00006\;(=0.006\%)$ (thin curves).
Considering at first the case $c_1=0$ at $T=0.8T_c(0)$
we find 
that the equilibrium values 
of the pressure, 
the chemical potential,
the liquid density,
and 
the vapor density 
are
$p_0\approx 0.00004118$,
$\mu_0\approx -0.00996108$,
$\rho_l\approx 0.30719568$,
and
$\rho_v\approx 0.02172324$,
respectively.
Assume further that in a system the concentration of excited atoms becomes $c_1=0.00006$.
For such a fluid the equilibrium values of the pressure is $p_0\approx 0.00003179$
and the vapor with the pressure $\approx 0.00004118$
becomes metastable with the value of the supersaturation parameter
$s\approx 1.29559738$.
Moreover,
the equilibrium values of the chemical potential, the liquid density, and the vapor density 
of the fluid with $c_1=0.00006$ at $T=0.8T_c(0)$ are
$\mu_0\approx -0.01049448$,
$\rho_l\approx 0.32745048$,
and
$\rho_v\approx 0.01596817$, 
respectively.

\begin{figure}[h]
\vspace{0cm}
\begin{center}
\includegraphics[width=7.5cm]{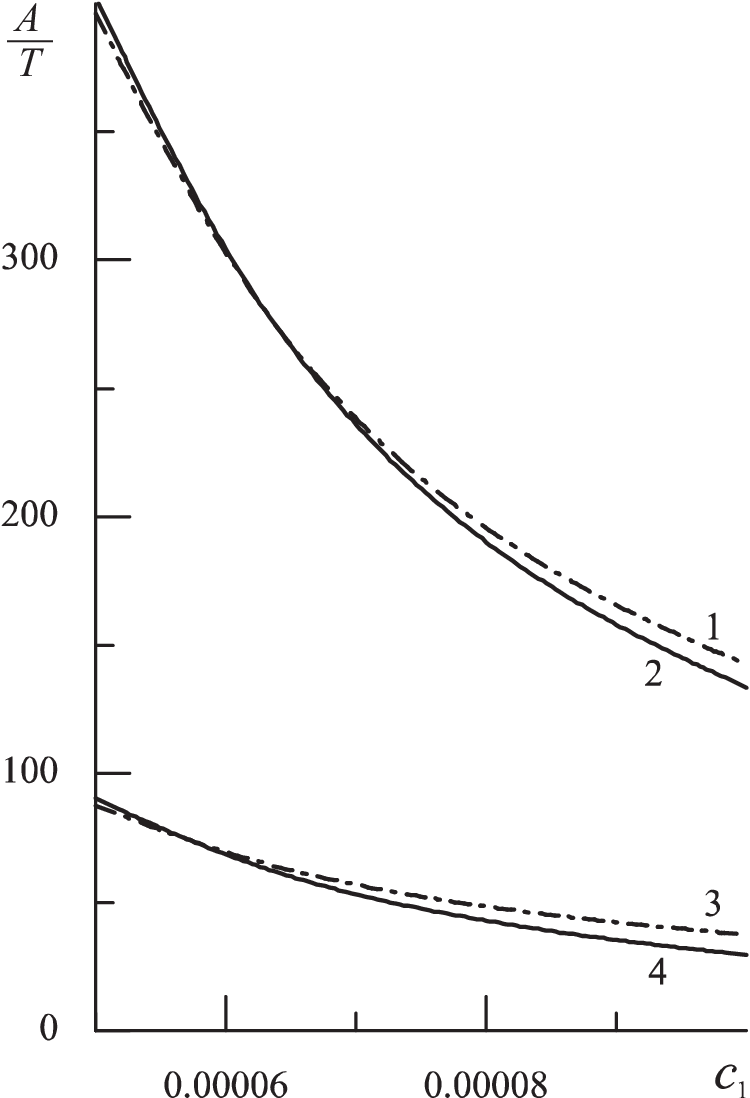}
\end{center}
\caption[]{
The dependence of the vapor-to-liquid nucleation barrier 
on the concentration of excited atoms $c_1$ at two temperatures 
$T=0.6T_c(0)$ (curves 1 and 2)
and
$T=0.8T_c(0)$ (curves 3 and 4).
The curves 1 and 3 were obtained using Eq. (\ref{01}),
the curves 2 and 4 were obtained using Eq. (\ref{02}).}
\label{fig2}
\end{figure}

To calculate the vapor-to-liquid nucleation barrier 
according to Eq. (\ref{01}) 
one has to find the surface tension $\gamma$.
Analyzing 
the density profile for a planar vapor-liquid interface 
(for this purpose we consider 
a two-phase system in a cylinder of the radius ${\cal{R}}$ and the height ${\cal{L}}$)
at $T=0.8T_c(0)$ and $c_1=0.00006$ 
and estimating $\Omega(T,\mu_0,V)$
we find according to the relation 
$\gamma\pi{\cal{R}}^2=\Omega(T,\mu_0,V)-(-p_0\pi{\cal{R}}^2{\cal{L}})$
the value of the surface tension
$\gamma=0.00051195$.
As a result one immediately gets the value of the vapor-to-liquid nucleation barrier
$\frac{A}{T}\approx 68.3294$
(see Fig. \ref{fig2}, dash-dotted curve 3).
Obviously, since $\frac{A}{T}$ becomes now finite 
(and decreases as $c_1$ increases)
the nucleation of liquid from vapor becomes now possible.

On the other hand, 
we can calculate the nucleation barrier on the basis of Eq. (\ref{02}).
First we estimate the Thompson radius 
$r^{\star}=\frac{2\gamma}{kT\rho_l\ln s}$
at $T=0.8T_c(0)$ and $c_1=0.00006$.
We obtain
$\frac{r^{\star}}{2}\approx 6.614283$.
We note that $r^{\star}$ is rather small
that may be a reason to go beyond the classical nucleation theory 
since the capillarity approximation cannot be justified for such small droplets.
Next we calculate the chemical potential for the supersaturated vapor with excited atoms
according to Eq. (\ref{10})
with $\rho\approx 0.02227178$
(this value of density follows from Eq. (\ref{11}) 
for $c_1=0.00006$, $T=0.8T_c(0)$ and $p\approx 0.00004118$),
$\mu=-0.01049448+0.00049852$,
and analyze the density profile of a spherical droplet in the supersaturated vapor 
seeking for a ``stable'' value of the grand thermodynamical potential 
which plays a role of $\Omega(T,\mu,V)$ in Eq. (\ref{02})
(for details see \cite{19}).
We find $\frac{A}{T}\approx 67.12$
that agrees with the value obtained within the frames of the classical nucleation theory.
The described calculations have to be repeated for other values of concentration $c_1$.
Moreover, 
we perform such calculations for several values of temperature.
Some of our findings are collected in Fig. \ref{fig2}.

The main conclusion which can be read off from Fig. \ref{fig2}
is as follows:
the vapor after appearance of excited atoms becomes metastable with $s>1$ 
and the nucleation barrier 
for vapor-to-liquid phase transition 
becomes essentially diminished.
This outcome agrees with a naive expectation 
that the long-range resonance dipole-dipole interactions 
should act in favor of liquid formation in vapor.
Although the present consideration permits 
to obtain the nucleation rate which can be measured experimentally
much more work is required to compare theory and experiment.
First,
we have to analyze in detail the results for nucleation rates \cite{15,16}
obtained using the upward thermal diffusion cloud chamber setup \cite{20}.
Second,
we should bare in mind
that the fluids whose photonucleation has been studied 
have more complicated particle structure and interparticle interactions.
Comparison with experiment can therefore be only qualitative at present,
and in this respect our results are consistent with the data reported in Refs. \cite{15,16}. 

\vspace{3mm}

One of the authors (O.~D.)
is grateful to the DAAD for the support of his visit 
to Philipps-Universit\"{a}t Marburg
in the autumn of 1995. 
He wishes to thank Dr.~Hermann~Uchtmann 
for kind hospitality and many stimulating conversations.

%
%
   \label{last@page}
   \end{document}